\def \VersionAuteur {}

\PassOptionsToPackage{svgnames,table}{xcolor}

\documentclass[a4paper,10pt]{llncs}

\makeatletter
\AtBeginDocument{%
  \@ifpackageloaded{hyperref}
  {\def\@doi#1{\href{https://doi.org/#1}
      {\ttfamily https://doi.org/#1}\egroup}}
  {\def\@doi#1{\ttfamily https://doi.org/#1\egroup}}
  \def\doi{\bgroup\catcode`\_=12\relax\@doi}}
\makeatother

\usepackage[utf8]{inputenc}
\usepackage[english]{babel}

\ifdefined \VersionAuteur
	\newcommand{\LongVersion}[1]{\ifdefined\VersionWithComments{\color{red!40!black}#1}\else#1\fi}
	\newcommand{\ShortVersion}[1]{\ifdefined\VersionWithComments{\color{black!40}#1}\fi}
\else
	\newcommand{\LongVersion}[1]{\ifdefined\VersionWithComments{\color{black!40}#1}\fi}
	\newcommand{\ShortVersion}[1]{\ifdefined\VersionWithComments{\color{red!40!black}#1}\else#1\fi}
\fi

\usepackage[ruled,vlined,linesnumbered]{algorithm2e}

\usepackage{subcaption}
\usepackage{float}
\usepackage{paralist} 
\usepackage{smartdiagram}
\usepackage{lscape}

\usepackage{amsmath} 
\usepackage{amssymb} 

\usepackage[misc,geometry]{ifsym} 

\usepackage{graphicx}
\graphicspath{{figures/}}

\LongVersion{

	\usepackage[backend=biber,backref=true,style=alphabetic,url=false,doi=true,defernumbers=true,sorting=anyt,maxnames=99]{biblatex} 
	\addbibresource{time4sys.bib}

	\renewbibmacro*{doi+eprint+url}{%
		\iftoggle{bbx:doi}
			{\color{black!40}\footnotesize\printfield{doi}}
			{}%
		\newunit\newblock
		\iftoggle{bbx:eprint}
			{\usebibmacro{eprint}}
			{}%
		\newunit\newblock
		\iftoggle{bbx:url}
			{\usebibmacro{url+urldate}}
			{}%
	}

}

\ifdefined\VersionWithComments
	\usepackage{draftwatermark}
	\SetWatermarkText{draft}
	\SetWatermarkScale{15}
	\SetWatermarkColor[gray]{0.9}
\fi

\definecolor{darkblue}{rgb}{0.0,0.0,0.6}
\definecolor{darkgreen}{rgb}{0, 0.5, 0}
\definecolor{darkpurple}{rgb}{0.7, 0, 0.7}
\definecolor{darkblue}{rgb}{0, 0, 0.7}

\usepackage[
		colorlinks=true,
	\ifdefined \VersionWithComments
		pagebackref=true,
	\fi
		citecolor=darkgreen,
		linkcolor=darkblue,
		urlcolor=darkpurple,
		pdfauthor={Étienne André, Jawher Jerray and Sahar Mhiri},%
		pdftitle={Time4sys2imi: A tool to formalize real-time system models under uncertainty},
	]{hyperref}

\usepackage[capitalise,english,nameinlink]{cleveref} 
\crefname{line}{\text{line}}{\text{lines}} 

\usepackage{tikz}
\usetikzlibrary{arrows,automata}
\tikzstyle{every node}=[initial text=]
\tikzstyle{location}=[rectangle, rounded corners, minimum size=12pt, draw=black, fill=blue!10, inner sep=2pt]
\tikzstyle{invariant}=[draw=black, dotted, inner sep=1pt] 

\newcommand{\rowHeader}{\rowcolor{blue!20}}

\definecolor{vertfonce}{rgb}{0.0, 0.5, 0.0}
\definecolor{rougefonce}{rgb}{1, 0.0, 0.0}

\newcommand{\cellYes}{\cellcolor{green!50}$\mathbf{\surd}$}
\newcommand{\cellNo}{\cellcolor{red!50}$\mathbf{\times}$}

\ifdefined\VersionWithComments
	\usepackage[colorinlistoftodos,textsize=footnotesize]{todonotes}
\else
	\usepackage[disable]{todonotes}
\fi
\newcommand{\gennote}[3]{\todo[linecolor=#2,backgroundcolor=#2!25,bordercolor=#2]{#3: #1}\xspace}
\newcommand{\ea}[1]{\gennote{#1}{blue}{ÉA}}
\newcommand{\jj}[1]{\gennote{#1}{orange}{JJ}}

\newcommand{\instructions}[1]{{\gennote{\bfseries #1}{red}{Instructions}}}

\usepackage{marginnote}
\ifdefined \VersionWithComments

	\newcommand{\todoinline}[1]{\mbox{}{\color{red}{\textbf{TODO}\ifx#1\\\else:\ \fi #1}}} 
\else
	\newcommand{\todoinline}[1]{}
\fi

\usepackage{verbatim} 

\newcommand{\cheddar}{Cheddar}

\newcommand{\imitator}{\textsf{IMITATOR}}

\newcommand{\outil}{\textsf{Time4sys2imi}}

\newcommand{\thales}{Thales Group}

\newcommand{\timeforsys}{Time4sys}

\newcommand{\uppaal}{\textsc{Uppaal}}

\ifdefined \VersionWithComments
 	\definecolor{colorok}{RGB}{80,80,150}
\else
	\definecolor{colorok}{RGB}{0,0,0}
\fi

\newcommand{\eg}{\textcolor{colorok}{e.\,g.,}\xspace}
\newcommand{\ie}{\textcolor{colorok}{i.\,e.,}\xspace}

\title{\outil{}:  A tool to formalize real-time system models under uncertainty\thanks{%
	\LongVersion{This is the author (and extended) version of the manuscript of the same name published in the proceedings of the 16th International Colloquium on Theoretical Aspects of Computing (\href{http://www.redcad.org/events/ictac2019/}{ICTAC 2019}).
	The final version is available at 
		\href{https://www.springer.com}{\nolinkurl{springer.com}}.
	}%
	This work is supported by the ASTREI project funded by the Paris Île-de-France Region,
	with the additional support of the ANR national research program PACS (ANR-14-CE28-0002)
	and
	ERATO HASUO Metamathematics for Systems Design Project (No.\ JPMJER1603), JST.
}}
\author{Étienne André\inst{1,2,3}\orcidID{0000-0001-8473-9555}, Jawher Jerray\inst{1}\orcidID{0000-0001-6170-7489}\Letter{} and Sahar Mhiri\inst{1}} 
\institute{Université Paris 13, LIPN, CNRS, UMR 7030, F-93430, Villetaneuse, France
\and%
JFLI, CNRS, Tokyo, Japan
\and%
National Institute of Informatics, Tokyo, Japan
}

\todo{This is the version with comments. To disable comments, comment out line~3 in the \LaTeX{} source.}

\begin{document}

\pagestyle{plain}

\maketitle

\setcounter{footnote}{0}

\thispagestyle{plain}

\todo{ajouter ref à TASE une fois paru !}

\begin{abstract}
\timeforsys{} is a formalism developed by \thales{}, realizing a graphical specification for real-time systems.
However, this formalism does not allow to perform formal analyses for real-time systems.
So a translation of this tool to a formalism equipped with a formal semantics is needed. 
We present here \outil{}, a tool translating \timeforsys{} models into parametric timed automata in the input language of \imitator{}.
This translation allows not only to check the schedulability of real-time systems, but also to infer some timing constraints (deadlines, offsets…)\ guaranteeing schedulability.
We successfully applied \outil{} to various examples. 
\end{abstract}


\instructions{ICTAC 2019: Short and tool papers should not exceed 10 pages.}

\ea{hello}
\jj{hello}

\section{Introduction}\label{section:introduction}


Due to the increasing complexity in real-time systems, designing and analyzing such systems is an important challenge, especially for safety-critical real-time systems, for which the correctness is crucial.
The scheduling problem for real-time systems consists in deciding which task the processor runs at each moment by taking into consideration the needs of urgency, importance and reactivity in the execution of the tasks.
Systems can feature one processor (``uniprocessor'') or several processors (``multiprocessor'').
Each processor features a scheduling policy, according to which it schedules new task instances.
Tasks are usually characterized by a best and worst case execution times (BCET and WCET), and are assigned a deadline and often a priority.
Tasks can be activated periodically (``periodic task''), sporadically (``sporadic tasks''), or be activated upon completion of another task---to which we refer to ``dependency'' or ``task chain''.
This latter feature is often harder to encode using traditional scheduling models.
Periodic tasks may be subject to a ``jitter'', \ie{} a variation in the period; all tasks can be subject to an ``offset'', \ie{} a constant time from the system start to the first activation of the task.
The schedulability problem consists in verifying that all tasks can finish their computation before their relative deadline, for a given scheduling policy.
This problem is a very delicate task: The origin of complexity arises from a large number of parameters to consider (BCET and WCET, tasks priorities, deadlines, periodic and sporadic tasks, tasks chains, etc.).
The schedulability problem becomes even more complicated when periods, deadlines or execution times become uncertain or completely unknown: we refer to this problem as schedulability under uncertainty.



\thales{}, a large multinational company specialized in aerospace, defense, transportation and security, developed a graphical formalism \timeforsys{}\footnote{%
	\url{https://github.com/polarsys/time4sys}
} to allow interoperability between timed verification tools.
\timeforsys{} responds to a need to unify the approaches within \thales{}: This formalism is being rolled out at TSA (Thales Airborne Systems) and studies are underway at TAS (Thales Alenia Space).
\timeforsys{} is now an open source framework, offering many features to represent real-time systems.
However, \timeforsys{} lacks for a formalization: it does not perform any verification nor simulation, nor can it assess the schedulability of the depicted systems.


Since \timeforsys{} does not allow to perform formal analyzes for real-time systems, a translation to a well-grounded formalism is needed to verify and analyze real-time systems.
In this paper, we present a tool \outil{} which allows to translate \timeforsys{} into parametric timed automata (PTAs)~\cite{AHV93} described in the input language of \imitator{}.
PTAs extend finite-state automata with clocks (\ie{} real-valued variables evolving at the same rate) and parameters (unknown timing constants).
PTAs are a formalism well-suited to verify systems where some timing delays are known with uncertainty, or completely unknown.
\imitator{}~\cite{AFKS12} is the \emph{de-facto} standard tool to analyze models represented using PTAs.
This translation allows not only to assess the schedulability of systems modeled using \timeforsys{}, but only to \emph{synthesize} some timing constants guaranteeing schedulability.

In~\cite{Andre19}, we presented a set of rules translating \timeforsys{} to PTAs.
We introduce here the tool performing this translation, with its practical description, as well as a set of case studies, absent from~\cite{Andre19}.


\paragraph{Related works}
Scheduling using (extensions) of timed automata was proposed in the past (\eg{} \cite{AAM06}).
For uniprocessor real-time systems only, (parametric) \emph{task automata} offer a more compact representation than (parametric) timed automata~\cite{FKPY07,NWY99,Andre17FMICS};
	however~\cite{FKPY07,NWY99} do not offer an automated translation and, while \cite{Andre17FMICS} comes with a script translating some parametric task automata to parametric timed automata, the case of multiprocessor is not addressed.
Schedulability analysis under uncertainty was also tackled in the past, \eg{} in \cite{CPR08,FLMS12,SSLAF13}.\ea{Todo after acceptance of this paper: add Ariane paper}
The main difference with our tool is that we allow here a systematic translation from an industrial formalism.

An export from \timeforsys{} is available to \cheddar{}~\cite{SLNM04}.
However, while \cheddar{} is able to deduce schedulability of real-time systems, it suffers from two main limitations:
\begin{enumerate}
	\item it does not allow task dependencies; and
	\item all timing constants must be fixed in order to study the schedulability.
\end{enumerate}
In contrast, our translation in \outil{} allows for both.

A model represented with \timeforsys{} can also be exported to MAST~\cite{GGPD01} which is an open-source suite of tools to perform schedulability analysis of real-time distributed systems.
However, the effectiveness of this tool is limited: it does not allow us to have a complete solution to our problem since it only works with instantiated systems, so we can not perform a real-time system with unknown parameters.

\paragraph{Outline}
\cref{section:time4sys} describes \timeforsys{}, and states the problem.
\cref{section:architecture} exposes the architecture of \outil{}. 
As a proof of concept, \cref{section:experiments} gives the results obtained on some examples.
We discuss future works in \cref{section:perspectives}.

\section{\timeforsys{} in a nutshell}\label{section:time4sys}

We review here \timeforsys{}, and make a few (minor) assumptions to ease our translation.

\timeforsys{} is a formalism that provides an environment to prepare the design phase of a system 
through the graphical visualization developed. \timeforsys{} contains two modes: Design and Analysis.
In our translation, we use the \timeforsys{} Design mode which uses a subset of the OMG MARTE standard~\cite{MARTE} as a basis for displaying a synthetic view to the real-time system.
This graphical representation encompasses all the elements and properties that can define a real-time system. 

The \timeforsys{} Design tool allows users to define the following elements:

\begin{itemize}
	\item \textbf{Hardware Resource}: a hardware resource in \timeforsys{} is a processor, and it contains a set of tasks; it is also assigned a scheduling policy.

	\item \textbf{Software Resource}: a software resource in \timeforsys{} is a task, and it features a (relative) deadline. 

	\item \textbf{Execution Step}: an execution step can be seen as a subtask. It is characterized by a BCET, a WCET, and a priority.
		In our translation, we assume that each software resource contains exactly one execution step.
		That is, we do not encompass for subtasks.

	\item \textbf{Event}: an event can be seen as an activation policy for tasks. There are two main types of Events:
		\begin{itemize}
			\item \textbf{PeriodicEvent}: defined by its period, its jitter and its phase (\ie{} offset).
			\item \textbf{SporadicEvent}: defined by its minimum and maximum interarrival times, its jitter and its phase.
		\end{itemize}

\end{itemize}

\begin{figure}[H]
		\centering
	\includegraphics[width=\textwidth]{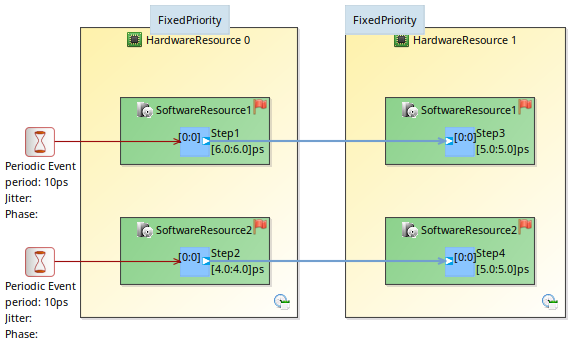}
	
	\caption{Example of a \timeforsys{} design}
	\label{fig:Example-Time4sys-Design}
\end{figure}

\begin{example}\label{example:running}
	\cref{fig:Example-Time4sys-Design} shows an example of a real-time system designed with \timeforsys{}.
	In this example, we have two hardware resources (HardwareResource0, HardwareResource1) both using fixed priority as a scheduling policy, two software resources (SoftwareResource1, SoftwareResource2) in each hardware resource, and four execution tasks, with the following timing constraints:
	\begin{itemize}
		\item Step1: $\mathit{WCET} = \mathit{BCET} = 6\,ps$
		\item Step2: $\mathit{WCET} = \mathit{BCET} = 4\,ps$
		\item Step3: $\mathit{WCET} = \mathit{BCET} = 5\,ps$
		\item Step4: $\mathit{WCET} = \mathit{BCET} = 5\,ps$
	\end{itemize}
	Finally, this example features two periodic events, both characterized by a 10\,ps period, a 0\,ps jitter and a 0\,ps phase (``offset'').
	\ea{il manque la priorité des tâches}

	In this example, we start executing with Step1 in the CPU HardwareResource0. After 6\,ps, the execution of Step1 ends so Step2 takes its place. At the same time, Step3 in the CPU HardwareResource1 starts performing. At $t=10\,ps$, the execution of Step2 finishes and a new period of Step1 starts, however at that time Step3 is still executing.
	So this real-time system is not schedulable
 \ie{} the period of StepT1 is strictly less than the WCET of Step1 plus the WCET of Step3.\ea{ça manque un peu d'explications, en fait}
\end{example}


\LongVersion{%
	\timeforsys{} Design can be used for different design modeling tool.
	It can be exported to different languages such as UML and AADL.
}

\subsection*{Objective}

The main objective of \outil{} is as follows:
given a real-time system with some unknown timing constants (period, jitter, deadlines…), \emph{synthesize} the timing constants for which the system is schedulable.
Note that, when all timing constants are known precisely, this problem is schedulability analysis.

\section{Architecture and principle}\label{section:architecture}

The main purpose of \outil{} is to perform the translation of \timeforsys{} models into the input language of \imitator{}.
The schedulability analysis itself is done by \imitator{}, using reachability synthesis.

\subsection{Targeted user}

The application is intended primarily for the designer of real-time systems, aiming to verify the schedulability of her/his system, or synthesize the timing constants ensuring schedulability.

\outil{} can automatically analyze a graphical representation of a real-time system realized by \timeforsys{}
using \imitator{}.
The end-user does not need to have skills on PTAs nor on model checking.

\begin{figure}[htbp!]
		\centering
	\includegraphics[scale=0.35]{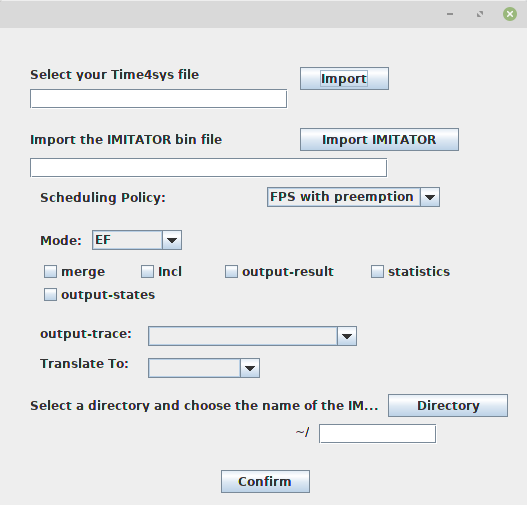}
	
	\caption{GUI of \outil{}}
	\label{fig:GUI}
\end{figure}

\outil{} allows the user to:
\begin{itemize}
\item Use the GUI of \outil{} (cf.~\cref{fig:GUI}) and configure the options of both the translation and \imitator{}.
\item Import an XML file generated by \timeforsys{}. This file contains the data that describes the real-time system to be analyzed.
\item Generate an \texttt{.imi} model analyzable by \imitator{}.
\end{itemize}


\subsection{User workflow}

The analysis of real-time systems, using the proposed translation, can be summed up in three main parts:
\begin{enumerate}
\item Graphical modeling of a real-time system containing all its components with \timeforsys{}. This part allows us to have a complete architecture of the system on the one hand. The architecture is encoded in an XML file automatically generated by \timeforsys{}. This file contains all the data needed to describe the system.
\item The second part 
	is the automatic translation of the XML file to the input language of \imitator{}, and is performed by \outil{}.
	\outil{} creates an \texttt{.imi} file that is analyzable with \imitator{}.
\item 	
	Finally, the user can run \imitator{} from \outil{} to get the answer to the schedulability problem.
	\ea{mais est-ce que tout est intégré, ou est-ce que l'utilisateur doit lui-même exécuter \imitator{} ?}
\end{enumerate}

The translation rules are described in~\cite{Andre19}.
In short, we translate each task, each task chain and each processor scheduling policy (earliest deadline first, rate monotonic, shortest job first…)\ into a PTA;
	most of these PTAs feature a special location corresponding to a deadline miss (\ie{} this location is reachable iff a deadline miss occurs).
Timing constants are encoded either as constants (if they are known) or as timing parameters (if they are unknown).
Then, we build (on-the-fly) the synchronous product of these PTAs.
Finally, the set of valuations for which the system is schedulable is exactly those for which the special deadline miss locations in the synchronous product are unreachable.
See \cite{Andre19} for details.

\subsection{Global architecture}

\ea{mais il manque des infos essentielles ici : quel est le langage de programmation de \outil{} ? Combien de lignes de code ? Quel système d'exploitation peut-on utiliser pour son utilisation ? Est-ce que tout est intégré ensemble (time4sys + traduction + \imitator{} ou faut-il faire plusieurs étapes ? etc.}

\outil{} is made of 5,500 lines of Java code, and 
can therefore run under any operating system. 
We explain in \cref{fig:architecture} the global architecture of the system.
\ea{ok, mais je vois pas bien les 3 parties sur la \cref{fig:architecture} ; on a l'impression que ta figure (qui est intéressante quand même) ne correspond pas bien au texte. }


\outil{} takes as input the \timeforsys{} model in XML, then we used the DOM parser to extract data.
	These data are translated into an abstract syntax for PTAs.
	We then translate these abstract PTAs into the concrete input language of \imitator{}.

\begin{figure}
\centering
\smartdiagramset{
                    border color=none,
                    back arrow disabled=true,
                }
\smartdiagram[flow diagram:horizontal]{\timeforsys{} model,
  Parsing, Translation, PTA model,\imitator{}}

	\caption{Workflow of \outil{}}
  \label{fig:architecture}
\end{figure}


%
%
%

\subsection{Detailed architecture}

\begin{figure}[tb]
	\centering
	\includegraphics[scale=0.55]{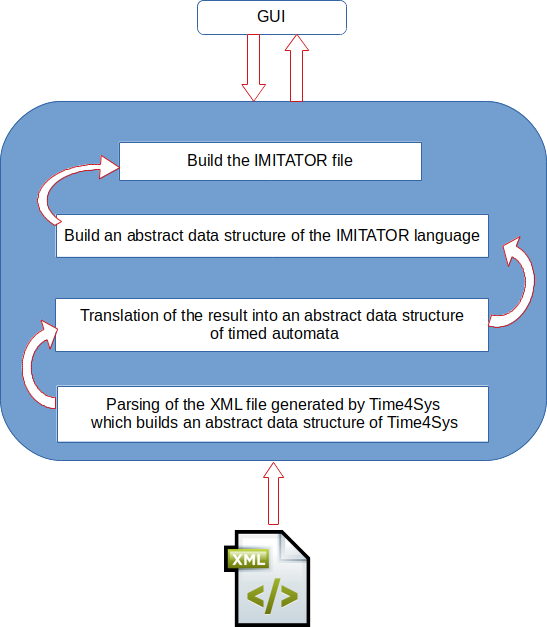}

	\caption{Detailed architecture}
	\label{fig:architecture_Tool}
\end{figure}

The global process is in \cref{fig:architecture_Tool}.

\begin{description}
	\item[Level 1] This level is the interface between the translation tool and the user: It allows the user to import the XML file to be translated, to choose the name of the \imitator{} model and to confirm the translation request.
\item[Level 2] This level is loaded by the translation of the XML file through the following steps:
\begin{enumerate}
\item Parsing the XML file that \timeforsys{} generates in order to get an abstract data structure from \timeforsys{}.
\item Translation of the result into an abstract data structure of PTAs.
\item Construct an \imitator{} file from the PTAs abstract data structure.
\end{enumerate}
\item[Level 3] This level shows the XML files generated by \timeforsys{} when designing a real-time system.

\end{description}

\section{Proof of concept}\label{section:experiments}

As a proof of concept to show the applicability of our translation tool, we 
modeled some real-time systems with \timeforsys{}, then we translated those models to PTAs using with \outil{} and analyzed them using \imitator{}.

We give in \cref{table:time-no-reac} a list of four case studies with, from top to bottom, the number of CPU, of tasks and task chains in the original \timeforsys{} model, followed by the number of automata, locations, clocks, discrete variables\footnote{%
	Discrete variables are global rational-valued variables that can be read and modified by the PTAs.
} and parameters in the translated \imitator{} target model.
We also give the name of the constants that are indeed parameterized (if any), and give the analysis time by \imitator{}.
The translation time using \outil{} is always negligible in our experiments.\ea{tu n'as pas un ordre de grandeur, tout de même ?}
Finally, we give whether the system is schedulable (if it is entirely non-parametric), or we give the condition for which it is schedulable.
The parametric results (\ie{} the constraints over the valuations ensuring schedulability) are given \LongVersion{in \cref{table:results} in \cref{appendix:results}}\ShortVersion{in~\cite{AJS19long}}.

\ea{dans \cref{table:time-no-reac}, j'aurais aimé voir le nombre d'automates, et de locations (si tu as gardé les données) ; tu peux ajouter ?}

We ran experiments on an ASUS X411UN Intel Core$\texttrademark$ i7-8550U 1.80\,GHz 
with 8\,GiB memory running Linux Mint~19 64\,bits.
All experiments were conducted using \imitator{} 2.10.4 ``Butter Jellyfish''.

Source, binaries, examples and results are available at \href{https://www.imitator.fr/static/ICTAC19}{\nolinkurl{www.imitator.fr/static/ICTAC19}}.
%

From \cref{table:time-no-reac}, we see that the analysis time using \imitator{} remains small, with the exception of the larger model with 11 concurrent tasks featuring dependencies, for which the analysis time using \imitator{} for a three-dimensional analysis becomes above 2~minutes.

\begin{example}
	Consider again the real-time system modeled in \cref{fig:Example-Time4sys-Design} using \timeforsys{}.
	We translate it using \outil{}\LongVersion{; the set of PTA obtained for this example are illustrated in \cref{fig:example3-pta} in \cref{appendix}}.
	
	First, we consider a non-parametric analysis: applying \imitator{} to the PTAs translated using \outil{} shows that the system is not schedulable, as it was expected from \cref{example:running}.
	
	Second, we parameterize the BCET and WCET of Step1.
	The result of the schedulability synthesis using \imitator{} yields the following constraint:
	 \(0 \leq \mathit{BCETStep1} \leq \mathit{WCETStep1} < 5 \).
	
	%
	This constraint explains why this real-time system was not schedulable when $\mathit{WCET} = \mathit{BCET} = 6$ \ie{} the values taken for $\mathit{WCET}$ and $\mathit{BCET}$ are not in the interval for which the system is schedulable.
	\end{example}

Additional examples with models and translated PTAs are given in \LongVersion{\cref{appendix}}\ShortVersion{\cite{AJS19long}}.

\begin{table}
	\centering
	\caption{Summary of experiments}
\scalebox{.8}{
\begin{tabular}{|c|c|c|c|c|c|c|c|c|}
   \hline
	\rowHeader{}
   \textbf{Case study} & \multicolumn{2}{c|}{Example 1[\cref{fig:Example-Time4sys-Design}]} & \multicolumn{2}{c|}{Example 2 [\LongVersion{\cref{fig:example1}}\ShortVersion{\cite{AJS19long}}]} & \multicolumn{2}{c|}{Example 3 [\LongVersion{\cref{fig:example2}}\ShortVersion{\cite{AJS19long}}]} & \multicolumn{2}{c|}{Example 4 [\LongVersion{\cref{fig:example4}}\ShortVersion{\cite{AJS19long}}]} \\
   \hline
   \hline
   \# CPU  & \multicolumn{2}{c|}{2} & \multicolumn{2}{c|}{1} & \multicolumn{2}{c|}{1} & \multicolumn{2}{c|}{4} \\
   \hline
   \# tasks & \multicolumn{2}{c|}{4} & \multicolumn{2}{c|}{4} & \multicolumn{2}{c|}{3} & \multicolumn{2}{c|}{11}\\
      \hline
	\# tasks chains & \multicolumn{2}{c|}{2} & \multicolumn{2}{c|}{0} & \multicolumn{2}{c|}{1} & \multicolumn{2}{c|}{4}\\   
	\hline
	\hline
	   \# number of automata & \multicolumn{2}{c|}{6} & \multicolumn{2}{c|}{9} & \multicolumn{2}{c|}{3} & \multicolumn{2}{c|}{12}\\
	\hline
	   \# total number of locations & \multicolumn{2}{c|}{22} & \multicolumn{2}{c|}{26} & \multicolumn{2}{c|}{14} & \multicolumn{2}{c|}{53}\\
    \hline
   \# clocks & \multicolumn{2}{c|}{8} & \multicolumn{2}{c|}{8} & \multicolumn{2}{c|}{6} & \multicolumn{2}{c|}{22}\\
      \hline
   \# discrete & \multicolumn{2}{c|}{4} & \multicolumn{2}{c|}{4} & \multicolumn{2}{c|}{3} & \multicolumn{2}{c|}{11}\\
	\hline
   \# parameters & 0 & 2 & 0 & 1 & 0 & 2 & 0 & 3\\
   \hline
   Parameters & - & WCETStep1  & - & DeadlineStep2& - & DeadlineStep1  & - & WCETStep5   \\
    &  &  BCETStep1 &  & &  &   &  &  BCETStep5  \\
    &  &   &  & &  &   &  &   DealineStep11 \\
    \hline
   Execution time (seconds) & 0.040 & 0.112 & 0.263 & 0.289 & 0.042 & 0.045 & 2.276 & 144.627\\
      \hline
   Schedulable? & \cellNo{} & \LongVersion{\hyperref[eq:condition1]{Condition1}} & \cellYes{} & \LongVersion{\hyperref[eq:condition2]{Condition2}} & \cellYes{} & \LongVersion{\hyperref[eq:condition3]{Condition3}} & \cellYes{} & \LongVersion{\hyperref[eq:condition4]{Condition4}} \\

   \hline
\end{tabular}
}
	\label{table:time-no-reac}
\end{table}

\section{Perspectives}\label{section:perspectives}


A short term future work will be to optimize our translation: while we followed the rules developed in~\cite{Andre19}, it is likely that varying the rules in order to optimize the size of the automata or reducing the clocks, may help to make the model more compact and the analysis more efficient.

Second, when the model is entirely non-parametric, we believe that using the \uppaal{} model checker~\cite{LPY97} instead of \imitator{} may be more efficient; for that purpose, we plan to develop a translator to the input language of \uppaal{} too; this implies to modify only the last step of our translation (from the abstract (P)TAs into the concrete input language of the target model checker).

Third, so far the analysis using \imitator{} is exact, \ie{} sound and complete; however, it may sometimes be interesting to get only \emph{some} ranges of parameter valuations for which the system is schedulable.
Such optimizations (on the \imitator{} side) should help to make the analysis faster.

Seeing from our experiments, it is unlikely that the toolkit made of \timeforsys{}, \outil{} and \imitator{} can analyze models with hundreds of processors and thousands of tasks, especially with unknown timing constants.
However, we believe that our approach can give first useful guarantees at the preliminary stage of system design and verification, notably to help designers to exhibit suitable ranges of timing parameters guaranteeing schedulability.
Finally, real-time systems with uncertain timing constants were recently proved useful when \thales{} published an open challenge\footnote{%
	``Formal Methods for Timing Verification Challenge'', in the WATERS workshop: \url{http://waters2015.inria.fr/challenge/}
} for a system (actually modeled using \timeforsys{}) with periods known with a limited precision only; while this problem was not strictly speaking a schedulability problem (but rather a computation of minimum/maximum execution times), it shed light on the practical need for methods to formally analyze real-time systems under uncertainty in the industry.

\section*{Acknowledgements}

We thank Romain Soulat and Laurent Rioux from Thales~R\&D for useful help concerning \timeforsys{}.

\LongVersion{
This work is partially supported by the ASTREI project funded by the Paris Île-de-France Region.

\begin{center}
	\includegraphics[width=.5\textwidth]{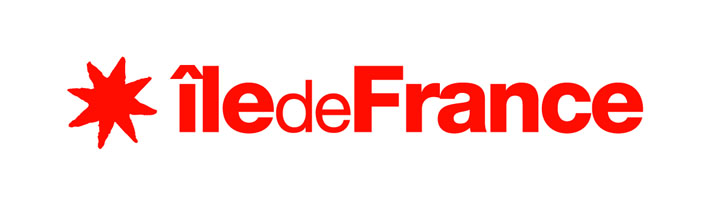}
\end{center}
}


	\newcommand{\CCIS}{Communications in Computer and Information Science}
	\newcommand{\ENTCS}{Electronic Notes in Theoretical Computer Science}
	\newcommand{\FI}{Fundamenta Informormaticae}
	\newcommand{\FMSD}{Formal Methods in System Design}
	\newcommand{\IJFCS}{International Journal of Foundations of Computer Science}
	\newcommand{\IJSSE}{International Journal of Secure Software Engineering}
	\newcommand{\IPL}{Information Processing Letters}
	\newcommand{\JLAP}{Journal of Logic and Algebraic Programming}
	\newcommand{\JLAMP}{Journal of Logical and Algebraic Methods in Programming} 
	\newcommand{\JLC}{Journal of Logic and Computation}
	\newcommand{\LMCS}{Logical Methods in Computer Science}
	\newcommand{\LNCS}{Lecture Notes in Computer Science}
	\newcommand{\RESS}{Reliability Engineering \& System Safety}
	\newcommand{\STTT}{International Journal on Software Tools for Technology Transfer}
	\newcommand{\TCS}{Theoretical Computer Science}
	\newcommand{\ToPNoC}{Transactions on Petri Nets and Other Models of Concurrency}
	\newcommand{\TSE}{IEEE Transactions on Software Engineering}
	
\LongVersion{

	\renewcommand*{\bibfont}{\small}
	\printbibliography[title={References}]
}
\ShortVersion{
	\bibliographystyle{splncs04} 
	\bibliography{time4sys}
}

\LongVersion{
\newpage
\appendix

\section{Additional details on the experiments}\label{appendix}

\subsection{Example without tasks chain}

We modeled an example with \timeforsys{} presented in \cref{fig:example1}. This example contains four periodic tasks without task chains.
\begin{figure}[H]
	\centering
  \includegraphics[ scale=0.6]{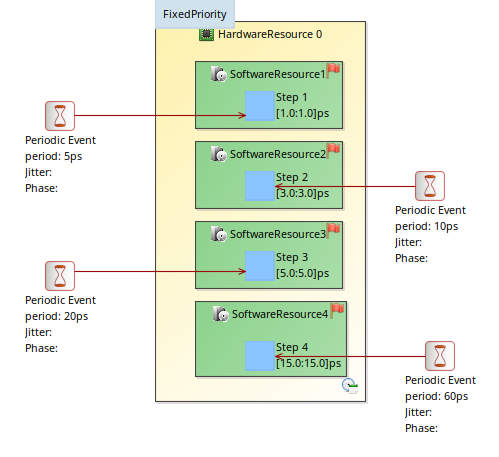}
  \caption{An example without tasks chain}
  \label{fig:example1}
\end{figure}

\cref{fig:example1-pta} illustrates the PTAs obtained from \cref{fig:example1} after the translation using \outil{}.
This graphics is automatically generated by \imitator{} (as the subsequent PTA depictions).

\begin{landscape}
\begin{figure}[H]
	\centering
  \includegraphics[scale=0.1]{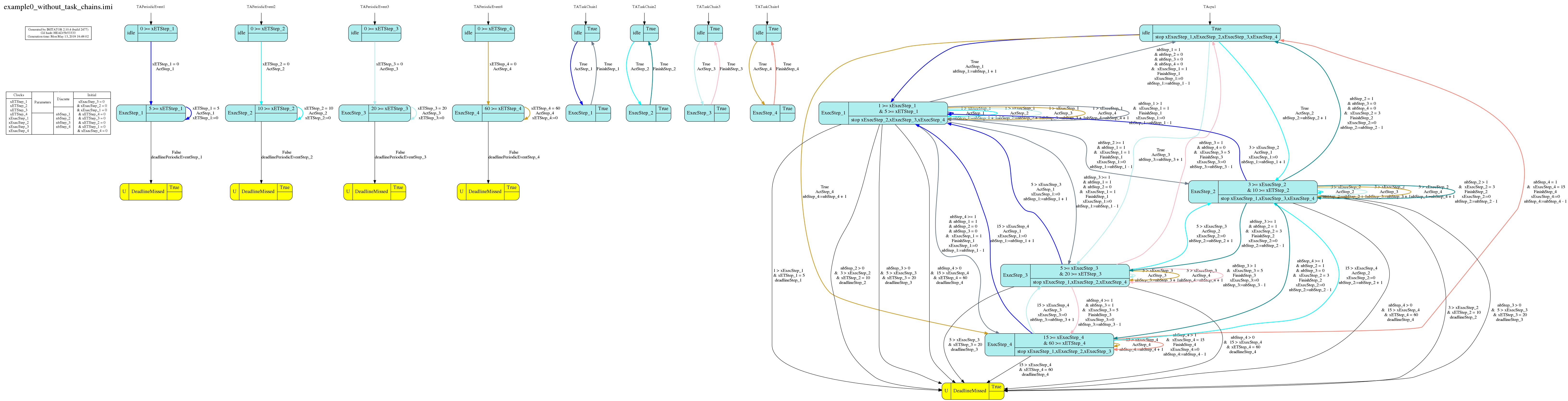}
  \caption{Translation of \cref{fig:example1}}
  \label{fig:example1-pta}
\end{figure}
  \end{landscape}
  
%

\subsection{Example with tasks chain}

We modeled an example with \timeforsys{} presented in \cref{fig:example2}.
This example contains three tasks, of which one is periodic; it contains also a tasks chain.
\begin{figure}[H]
	\centering
  \includegraphics[ scale=0.45]{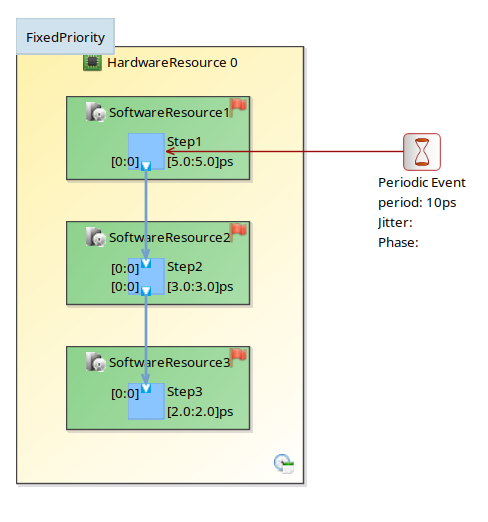}
  \caption{An example with tasks chain}
  \label{fig:example2}
  \end{figure}

\cref{fig:example2-pta} illustrates the PTAs obtained from \cref{fig:example2} after the translation using \outil{}.

  \begin{landscape}
  \begin{figure}[H]
	\centering
  \includegraphics[ scale=0.18]{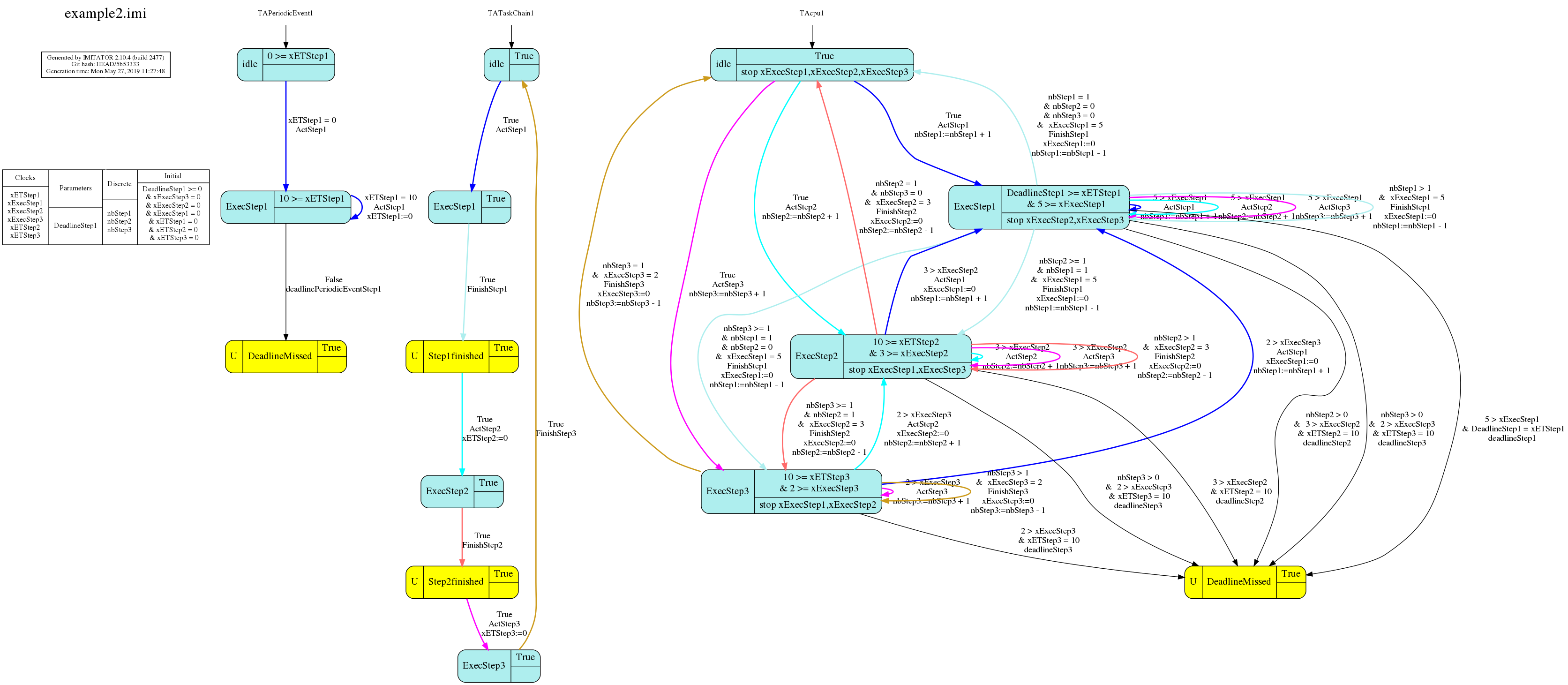}
  \caption{Translation of \cref{fig:example2}}
  \label{fig:example2-pta}
\end{figure}
\end{landscape}

\subsection{Example with two CPU}

Consider again the real-time system modeled in \cref{fig:Example-Time4sys-Design}.\\
  
  \cref{fig:example3-pta} illustrates the PTAs obtained from \cref{fig:Example-Time4sys-Design} after the translation using \outil{}.

  \begin{landscape}
    \begin{figure}[H]
	\centering
  \includegraphics[ scale=0.15]{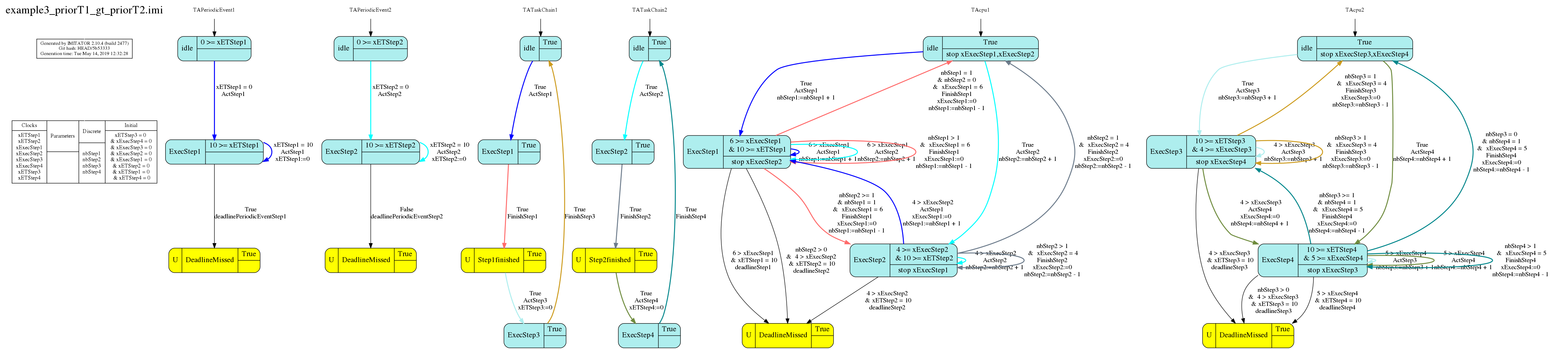}
  \caption{Translation of \cref{fig:Example-Time4sys-Design}}
  \label{fig:example3-pta}
\end{figure}
  \end{landscape}

\subsection{Example with 4 CPUs and 11 tasks}

We modeled an example with \timeforsys{} presented in \cref{fig:example4}. This example contains four CPU and eleven tasks; it also contains four tasks chains.
\begin{landscape}
\begin{figure}[H]
	\centering
  \includegraphics[ scale=0.35]{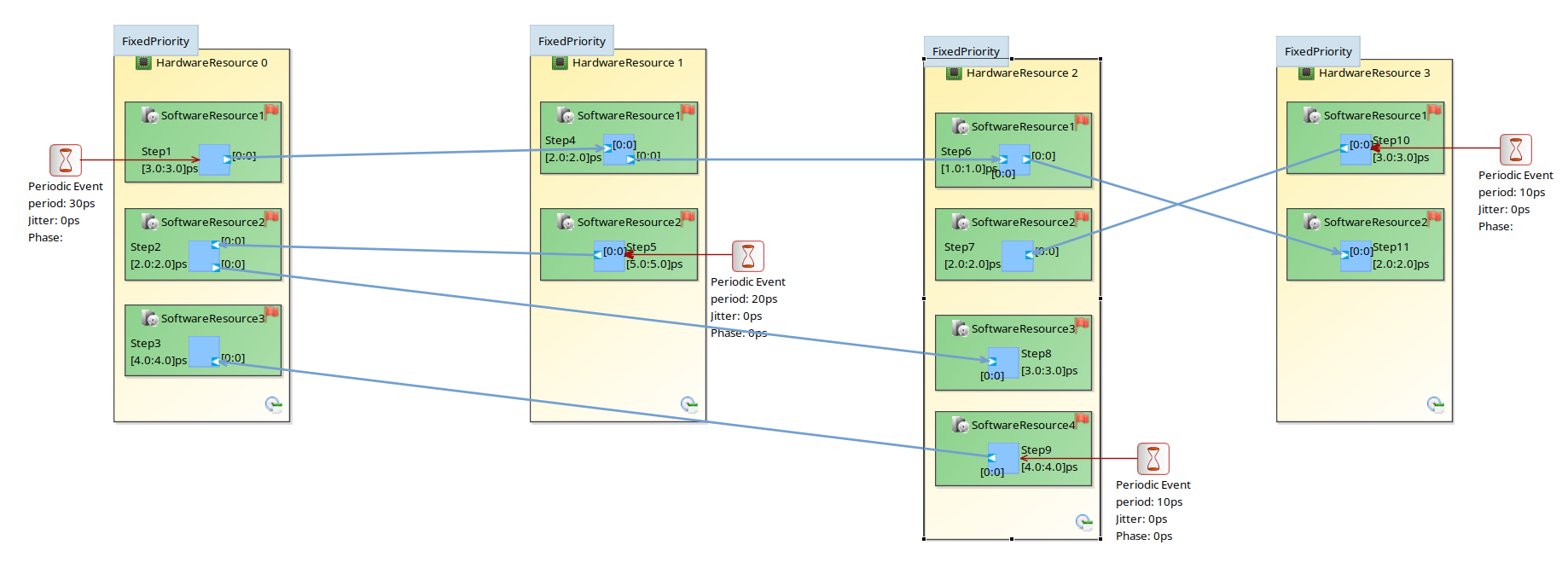}
  \caption{General example}
  \label{fig:example4}
  \end{figure}
  \cref{fig:example4-pta} illustrates the PTAs obtained from \cref{fig:example4} after the translation using \outil{}

      \begin{figure}[H]
	\centering
  \includegraphics[ scale=0.1]{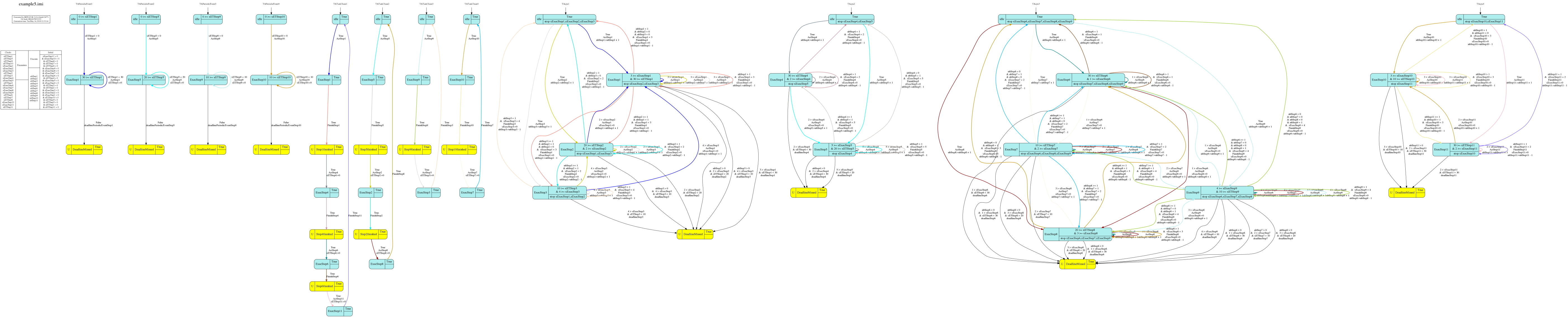}
  \caption{Translation of \cref{fig:example4}}
  \label{fig:example4-pta}
\end{figure}

  \end{landscape}

\subsection{Results}\label{appendix:results}

\begin{table}[H]
	\centering
	\caption{Synthesized constraints}
\begin{tabular}{|c|c|c|c|}

\hline
\rowHeader{}
Condition1\label{eq:condition1} & 
Condition2\label{eq:condition2} & 
Condition3\label{eq:condition3} & 
Condition4\label{eq:condition4}\\
\hline
   5 $>$ WCETStep1 & DeadlineStep2 $>=$ 4 & DeadlineStep1 $>=$ 5 & WCETStep5 $>=$ BCETStep5\\
   
   \& BCETStep1 $>=$ 0 & & & \& BCETStep5 $>=$ 0\\
   
   \& WCETStep1 $>=$ BCETStep1 & & & \& 15 $>$ WCETStep5 \\
  
 & & & \& DeadlineStep11 $>=$ 5 \\
 & & &  \textbf{OR}\\
   & & &  5 $>$ DeadlineStep11\\
 & & &  \& BCETStep5 $>$ 4\\
  & & & \& DeadlineStep11 $>=$ 2\\
 & & &  \& WCETStep5 $>=$ BCETStep5\\
  & & & \& 6 $>=$ WCETStep5\\
  & & & \textbf{OR}\\
   & & &  DeadlineStep11 $>=$ 2\\
  & & & \& 5 $>$ DeadlineStep11\\
 & & &  \& 15 $>$ WCETStep5\\
  & & & \& BCETStep5 $>$ 14\\
  & & & \& WCETStep5 $>=$ BCETStep5\\
 \hline
 \end{tabular}
	\label{table:results}
\end{table}

}

\end{document}